\documentclass[%
 reprint,
 amsmath,amssymb,
 aps,
pre,
floatfix,
]{revtex4-1}

\usepackage{graphicx}
\usepackage{dcolumn}
\usepackage{bm}
\usepackage{epstopdf}
\usepackage{siunitx}
\usepackage{svg}
\usepackage{gensymb}
\usepackage{float} 

\begin{document}


\title{Current-Driven Domain Wall Motion in Curved Ferrimagnetic Strips Above and Below the Angular Momentum Compensation}

\author{D. Osuna Ruiz *\textsuperscript{1}} 
\email{osunaruiz.david@usal.es}
\author{O. Alejos\textsuperscript{2}}
\author{V. Raposo\textsuperscript{1}}
\author{E. Martínez\textsuperscript{1}}%
 
\affiliation{%
\textsuperscript{1}Department of Applied Physics, University of Salamanca, Salamanca 37008, Spain
 }%
\affiliation{%
\textsuperscript{2}Department of Electricity and Electronics, University of Valladolid, Valladolid, Spain.
 }%
\date{\today}

\begin{abstract}
Current driven domain wall motion in curved Heavy Metal/Ferrimagnetic/Oxide multilayer strips is investigated using systematic micromagnetic simulations which account for spin-orbit coupling phenomena. Domain wall velocity and characteristic relaxation times are studied as functions of the geometry, curvature and width of the strip, at and out of the angular momentum compensation. Results show that domain walls can propagate faster and without a significant distortion in such strips in contrast to their ferromagnetic counterparts. Using an artificial system based on a straight strip with an equivalent current density distribution, we can discern its influence on the wall terminal velocity, as part of a more general geometrical influence due to the curved shape. Curved and narrow ferrimagnetic strips are promising candidates for designing high speed and fast response spintronic circuitry based on current-driven domain wall motion. 

\end{abstract}

\pacs{Valid PACS appear here}
\maketitle


\section{\label{sec:level1}introduction}

A magnetic domain wall (DW) is the transition region that separates two uniformly magnetized domains \cite{hubert98}. These magnetic configurations are interesting due to fundamental physics, but also due to potential technological applications \cite{Parkinpat,parkinracetrack}. In fact, during the last decades DWs have been at the core of theoretical and experimental studies which have  provided with a deep understanding of different spin-orbit coupling phenomena \cite{miron,emori,haazen,ryu,inertia}. For instance, straight stacks where an ultra-thin ferromagnetic (FM) layer is sandwiched between a heavy metal (HM) and an oxide (Ox), present perpendicular magnetic anisotropy (PMA) and therefore, the domains are magnetized along the out-of-plane direction of the stacks: \textit{up} ($+\vec{u}_z$) or \textit{down} ($-\vec{u}_z$). DWs in these HM/FM/Ox stacks adopt an homochiral configuration due to the Dzyaloshinskii-Moriya interaction (DMI) \cite{DZYALOSHINSKY1958241,emori,ryu}. Adjacent DWs have internal magnetic moments along the longitudinal direction ($\vec{m}_{DW}=\pm \vec{u}_{x}$), and the sense is imposed by the sign of the DMI, which in turns depends on the HM \cite{emori}. For left-handed stacks such as Pt/Co/AlO, \textit{up-down} (UD) and \textit{down-up} (DU) DWs have internal moments with $\vec{m}_{DW}=-\vec{u}_x$ and $\vec{m}_{DW}=+\vec{u}_x$, respectively \cite{emori}. These DWs are driven with high efficiency by injecting electrical currents along the longitudinal direction of the HM/FM/Ox stack \cite{miron}. Due to the spin-Hall effect \cite{emori}, the electrical current in the HM generates a spin polarized current which exerts spin-orbit torques (SOTs) on the magnetization of the FM layer, and drives series of homochiral DWs which are displaced along the longitudinal direction ($x$-axis). DW velocities of $V_{DW}\sim 500$ m/s have been reported upon injection of current densities of $J_{HM}\sim 1$ TA/m$^2$ in Pt/Co/AlO \cite{miron}. Consequently, these stacks have been proposed to develop highly-packed magnetic recording devices, where the information coded in the domains between DWs can be efficiently driven by pure electrical means. Both UD and DU DWs move with the same velocity along straight stacks, but some implementations of these memory or logic devices would require to design 2D circuits, where straight parts of HM/FM/Ox stack are connected each other with curved or semi-rings sections. However, recent experimental observations \cite{garg} and theoretical studies \cite{Alejos} have pointed out that adjacent UD and DU DWs move with different velocity along curved HM/FM/Ox stacks, which is detrimental for applications because the size of the domain between adjacent DWs changes during the motion, with the perturbation of the information coded therein. Therefore, other systems must be proposed in order to design reliable 2D circuits for DW-based memory and logic devices.

\par Other stacks with materials and/or layers with antiferromagnetic coupling, such as synthetic antiferromagnets (SAF) and ferrimagnetic (FiM), have proven to outperform FM in terms of current-driven DW dynamics \cite{yang,Alejos,PhysRevLett.121.057701_Liu_Siddiqui,Caretta2018FastCD,MARTINEZ2019165545}. Ultrafast magnetization dynamics in the THz regime, marginal stray field effects and insensitivity to external magnetic fields are other significant advantages of materials with antiferromagnetic coupling with respect to their FM counterparts. As conventional antiferromagnets (AFs), FiM alloys are also constituted by two specimens, typically a rare earth (RE) and transition metal (TM), that form two ferromagnetic sublattices antiferromagnetically coupled to each other. GdFeCo, GdFe or TbCo are archetypal FiM alloys, with the RE being Gd or Tb and the TM being FeCo or Co. In contrast to AFs with zero net magnetization, the magnetic properties of FiMs, such as magnetization and coercivity, are largely influenced by the relative RE and TM composition (or equivalently, temperature). This fact offers additional degrees of freedom to control the current-driven DW velocity. The spontaneous magnetization of each sublattice $M_{S,i}$ can be tuned by changing the composition of the FiM and/or the temperature of the ambient ($T$) \cite{PhysRevLett.121.057701_Liu_Siddiqui,Caretta2018FastCD}. 
For a given composition of the FiM (RE$_x$TM$_{1-x}$), there are two relevant temperatures below the Curie threshold. One is the magnetization compensation temperature ($T_M$) at which the saturation magnetization of the two sublattices are equal ($M_{s1}\left(T_M\right)=M_{s2}(T_M)$), so the FiM behaves as a perfect antiferromagnetic material, with zero net magnetization and diverging coercive field. The other is the temperature at which the angular momentum compensates, $T_A$, at which $M_{s1}\left(T_A\right)/\gamma_1=M_{s2}(T_A)/\gamma_2$, where $\gamma_i$ is the gyromagnetic ratio of each sublattice ($i$:1,2 for 1:TM and 2:RE). As the gyromagnetic ratio depends on the Landé factors ($g_i$) which are different for each sublattice, the angular compensation temperature $T_A$ is in general different from the magnetization compensation temperature ($T_M$). Consequently, the FiM have a net magnetization at $T_A$, so conventional techniques used for FMs can be also adopted to detect the magnetic state of FiM samples \cite{arpaci2021}. Moreover, recent experimental observations have evidenced that the current-driven DW velocity along straight HM/FiM stacks can be significantly optimized at the angular momentum compensation temperature ($T=T_A$), with velocities reaching $V_{DW}\sim2000$ m/s for typical injected density current of $J_{HM}\sim1$ TA/m$^2$ along the HM underneath \cite{Caretta2018FastCD}. The DW velocity drops either below ($T<T_A$) and above ($T>T_A$) angular momentum compensation. Note that alternatively to tuning the temperature for a fixed composition $x$ of the FiM alloy RE$_x$TM$_{1-x}$, even working at room temperature ($T=300$ K) the DW velocity peaks at a given composition where angular momentum compensates \cite{PhysRevLett.121.057701_Liu_Siddiqui}. Therefore, both studies, either fixing the composition ($x$) and changing temperature of the ambient ($T$), or fixing the ambient temperature and modifying the FiM composition are equivalent for our purposes of DW dynamic. Although the current-driven DW motion (CDDWM) along HM/FiM stacks suggests their potential for memory and logic applications, previous studies have been mainly focused on straight FiM strips \cite{Caretta2018FastCD,PhysRevLett.121.057701_Liu_Siddiqui,MARTINEZ2019165545}. The further develop of novel DW-based devices also requires to analyze the dynamics of DWs along HM/FiM with curved parts which would connect straight paths to design any 2D circuit. Such investigation of the dynamics along curved is still missing, and it is the aim of the present study.

\par Here we theoretically explore the CDDWM along curved HM/FiM stacks by means of micromagnetic ($\mu$m) simulations. Our modeling allows us to account for the magnetization dynamics in the two sublattices independently. We explore the CDDWM below, at and above the angular momentum compensation (AMC) for different curved samples, with different widths and curvatures, and considering the realistic spatial distribution of the injected current along the HM. In particular, we will infer and isolate the relevance of different aspects governing such dynamics, as the role of the non-uniform current and other purely geometrical aspects of the curved shape. This work completes previous studies on straight samples \cite{PhysRevLett.121.057701_Liu_Siddiqui,Caretta2018FastCD,MARTINEZ2019165545,Alejos}, and will be practical for designing more compact and efficient DW-based devices. The rest of the paper is organized as follows. In Sec. II we describe the numerical details of the micromagnetic model along with the material parameters and the geometrical details of the evaluated samples. Sec. III presents the micromagnetic results of the CDDWM in different scenarios. Firstly, exploring the role of the FiM sample width ($w$) for a fixed the curvature ($\rho$, given by the inverse of the average radius, $\rho=1/r_e$), and secondly fixing the width and varying the curvature. After that, we present results which allow us to infer the role of non-uniform current and geometrical aspect ($w,\rho$) comparing curved and straight samples. The main conclusions are summarized in Sec. IV.

\section{\label{sec:level2}Materials and methods}

\par CDDWM is numerically studied here along curved HM/FiM stacks as schematically shown in Fig. 1, where $r_i$ and $r_o$ are the inner and outer radius $r_o$ respectively, and $r_e= (r_o+\ r_i)/2$ is the mean effective radius. $w$ and $ t_{FiM}$ are the width and the thickness of the FiM respectively. The relaxed magnetization configuration of the sublattice $i=1$, shown in Fig. 1 (opposite configuration in sublattice $i=2$), serves as the initial state to study the CDDWM upon of current injection along the HM underneath. The temporal evolution of the magnetization of each sublattice is given by the Landau-Lifshitz-Gilbert equation (LLG) \cite{LANDAU199251}, 
\begin{gather}
\frac{d\vec{m}_{i}(t)}{dt} = -\gamma_{0,i}\vec{m}_{i}(t)\times \vec{H}_{eff,i} + \alpha_i \vec{m}_{i}\times \frac{d\vec{m}_{i}(t)}{dt} + \vec{\tau}_{SOT,i}, 
\end{gather}
where here the sub-index $i$ stands for $i$: 1 and 2 sublattices respectively. $\gamma_i=g_i\mu_B/\hbar$  and $\alpha_i$ are the gyromagnetic ratios and the Gilbert damping constants, respectively. $g_i$ is the Landé factor of each layer, and ${\vec{m}}_i\left(\vec{r},t\right)={\vec{M}}_i/M_{s,i}$ is the normalized local magnetization to its saturation value ($M_{s,i}$), defined differently for each sublattice: $M_{s,i} (i:1,2)$. In our micromagnetic model the FiM strip is formed by computational elementary cells, and within each cell we have two magnetic moments, one for each component of the FiM. The respective effective field (${\vec{H}}_{eff,i}$) acts on the local magnetization of each sublattice (${\vec{m}}_i\left(\vec{r},t\right)$), and it is the sum of the magnetostatic, the anisotropy (PMA), the DMI and the exchange fields \cite{Alejos,MARTINEZ2019165545}. The magnetostatic field on each local moment in the sublattice is numerically computed from the average magnetization of each elementary cell using similar numerical techniques as for the single FM case (see \cite{Alejos,MARTINEZ2019165545}). 
We checked that the demagnetising field has a marginal influence in the simulation results compared to other contributions to the effective field.
For the PMA field, the easy axis is along the out-of-plane direction ($z$-axis), and the anisotropy constants for each sublattice are $K_{u,i}$ (PMA constant). $D_i$ is the DMI parameter for each sublattice $i: 1,2$ \cite{Alejos,MARTINEZ2019165545}. The exchange field of each sublattice includes the interaction with itself (intra-lattice exchange interaction, ${\vec{H}}_{exch,i}$) and with the other sublattice (inter-lattice exchange interaction, ${\vec{H}}_{exch,12}$). The inter-lattice exchange effective field is computed as for a single FM sample, $H_{exch,i}=\frac{2A_i}{\mu_0M_{s,i}}\nabla^2{\vec{m}}_i$, where $A_i$ is the intralattice exchange parameter. The inter-lattice exchange contribution ${\vec{H}}_{exch,12}$ to the effective field ${\vec{H}}_{eff,i}$, acting on each sublattice is computed from the corresponding energy density, $\omega_{exch,i}=-B_{ij}{\vec{m}}_i\cdot{\vec{m}}_j$, where $B_{ij}$ (in $[\text{J}\ \text{m}^{-3}]$) is a parameter describing the inter-lattice exchange coupling between sublattices (here, we used the notation $i:1$ and $j:2$). 

\par In Eq.(1), ${\vec{\tau}}_{SOT,i}$ are the SOTs acting on each sublattice, which are related to the electrical current along the HM (${\vec{J}}_{HM}$). Based on preliminary studies \cite{martinez2014}, here we assume that ${\vec{\tau}}_{SOT,i}$ is dominated by the spin Hall effect (SHE), so ${\vec{\tau}}_{SOT,i}=-\gamma_0H_{SL}{\vec{m}}_i\times\left({\vec{m}}_i\times\vec{\sigma}\right)$ where $H_{SL}=\frac{\hbar\theta_{SH,i}J_{HM}}{2\left|e\right|\mu_0M_st_{FiM}}$ \cite{slonczewski1996current}. $\hbar$ is the Planck constant, and $\theta_{SH,i}$ is the spin Hall angle, which determines the ratio between the electric current and the spin current ($J_s=\theta_{SH}J_{HM}$) for each sublattice. $\vec{\sigma}={\vec{u}}_J\times{\vec{u}}_z$ is the unit vector along the polarization direction of the spin current generated by the SHE in the HM, being orthogonal to both the direction of the electric current ${\vec{u}}_J$ and the vector ${\vec{u}}_z$ standing for the normal to the HM/FiM interface. For a longitudinal current (${\vec{u}}_J={\vec{u}}_x$), the spin current is polarized along the transverse direction, $\vec{\sigma}=-{\vec{u}}_y$. For curved samples where the current density ${\vec{J}}_{HM}=J_{HM}\left(r\right){\vec{u}}_J$ has azimuthal direction (${\vec{u}}_J={-\vec{u}}_\phi)$, the direction of the polarization is radial, $\vec{\sigma}={\vec{u}}_J\times{\vec{u}}_z={\vec{u}}_r$, as shown in Fig 1. A potential difference is applied between the ends of the curved track to inject current in the right circulation. Therefore, a gap of 25 nm is also modelled, leading to a split ring shape for the strip (see inset in Fig. 1). The spatial distribution of current as a function of the radial coordinate ($r_i<r<r_o$) is taken from \cite{model2018,Alejos}, and it depends on the width ($w$) and the radial distance ($r$) as $J_{HM}\left(r\right)=J_0w/(r\log{\left(1+w/r_i\right)})$, where $J_0$ is the nominal, uniform current density, in an equivalent straight strip of same cross-section ($w\times t_{HM}$, where $t_{HM}$ is the thickness of the HM strip). 

\par In order to illustrate the current-driven DW dynamics along curved HM/FiM stacks we fix $t_{FiM}=6$ nm, and samples with different widths ($w$) and radii ($r_e$) were evaluated. The following common material parameters were adopted for the two sublattices $i:1,2$: $A_i=70$ pJ/m, $K_{u,i}=1.4\times{10}^6$ J/m$^3$, $\alpha_i=0.02$, $D_i=0.12$ J/m$^2$, $\theta_{SH,i}=0.155$. The strength of the antiferromagnetic coupling between the sublattices was fixed to $B_{ij}\equiv B_{12}=-0.9\times{10}^7$ J/m$^3$. The gyromagnetic ratios ($\gamma_i=g_i\mu_B/\hbar$) are different due to the different Landé factor: $g_1=2.05$ and $g_2=2.0$. The saturation magnetization of each sublattice $M_{s,i}$ can be tuned with the composition of the FiM and/or with the temperature of the ambient ($T$). Here, we assume the following temperature dependences for each sublattice: $M_{s,i}\left(T\right)=M_{s,i}(0)\left(1-\frac{T}{T_C}\right)^{a_i}$, where $T_C=450$ K is the Curie temperature of the FiM, $M_{s,1}\left(0\right)=1.4\times{10}^6$ A/m and $M_{s,2}\left(0\right)=1.71\times{10}^6$ A/m are the saturation magnetization at zero temperature, and $a_1=0.5$ and $a_2=0.76$ are the exponents describing the temperature dependence of the saturation magnetization of each sublattice. The temperature at which the net saturation magnetization vanishes ($M_{s,1}(T_M)=M_{s,2}(T_M)$) is $T_M=241.5$ K, and the angular momentum compensation temperature corresponding to $M_{s,1}(T_A)/g_1=M_{s,2}(T_A)/g_2$, is $T_A=260$ K. We evaluate the CDDWM below, at and above the angular momentum compensation adopting three representative temperatures: $T=220$ K $<T_A$, $T=260$ K $=T_A$ and $T=300$ K $>T_A$. Samples were discretized using a 2D finite difference scheme using computational cells with $\Delta x=\Delta y=0.2$ nm and $\Delta z=t_{FiM}$. Several tests were carried to certify that the presented results are free of discretization errors. 

\begin{figure}[ht]
\centering 
\includegraphics[width=8.5cm]{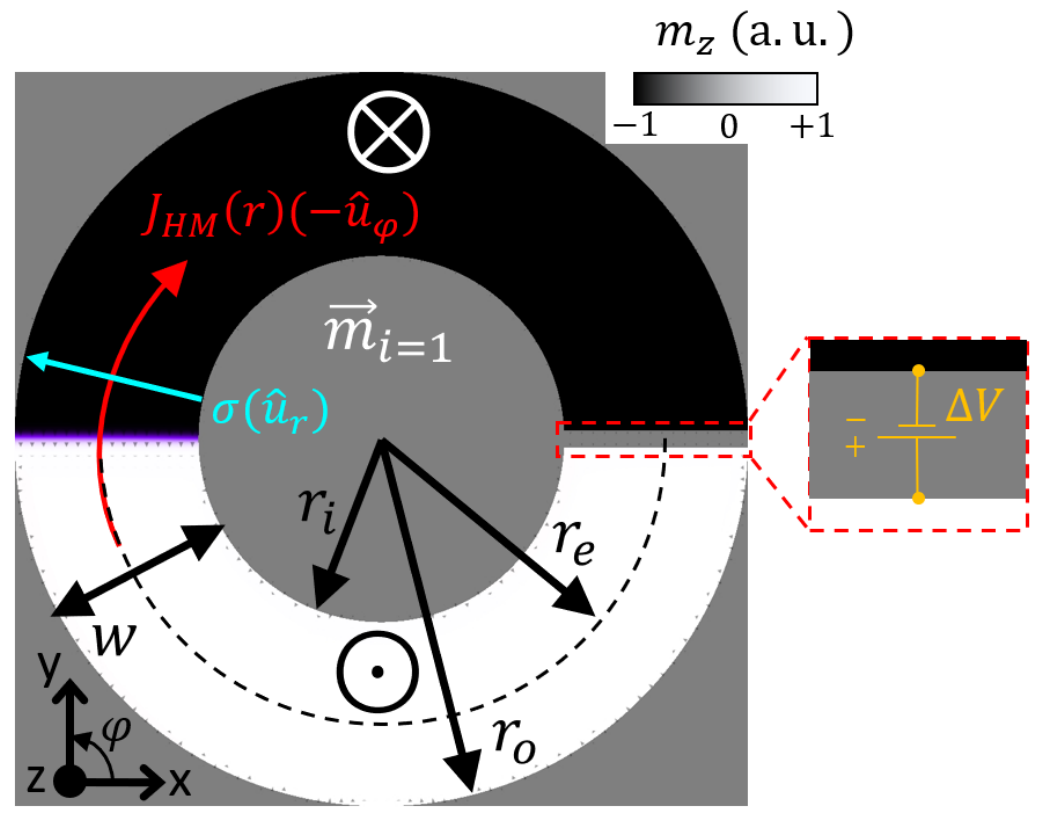}
\caption{Scheme showing the relaxed states of spins in sub-lattice $i=1$, in the positive z-direction (white domain), in the negative z-direction (black domain) and in the plane of the strip for an \lq Up to Down' (UD) domain wall (purple) according to the current direction, for an exemplary curved strip. The direction of the applied electric current (red arrow) in the Heavy Metal beneath the magnetic strip, generated from a potential difference $\Delta V$ (see inset), is shown as well as the geometrical parameters of the strip.}  \label{teardrop_init}
\end{figure}

\section{Micromagnetic results}

Due to the several combination of parameters to consider in our study, we divided this section in three sub-sections: (A) The study on the influence of the strip width ($w$); (B) the study on curvature ($\rho = r_e^{-1}$); and (C) same studies for a straight strip with identical material parameters, to explore by comparison the effects of curvature on the DW dynamics.  In parts (A) and (B), scenarios for three different temperatures, $T_{1}=220$ K$, T_{2}=260 $ K$, T_{3}=300$ K are considered, to study the DW motion below the AMC ($T_1$), at the AMC ($T_2$) and above the AMC ($T_3$). We also define and refer to $T_3 = 300$ K as for \lq room temperature' in our study. Note that a change in temperature only affects $M_S$ in our model, therefore it has equivalent effects to changing material composition \cite{PhysRevLett.121.057701_Liu_Siddiqui}. In addition to DW velocity, we also characterize the inertial motion of the DW as a function of current density. As an example, Fig. 2 shows typical results of the DW position and its velocity in a ring-like strip ($w= 256$ nm and $r_e=384$ nm) under a density current $J_{HM}=2$ TA/m$^2$ and at $T=260$ K. Qualitatively similar results are obtained at $T=220$ K and $T=300$ K (not shown). Insets show the (clockwise) DW displacement as a function of time for one sublattice $(i=1)$.

\begin{figure}[t]
\centering 
\includegraphics[width=8cm]{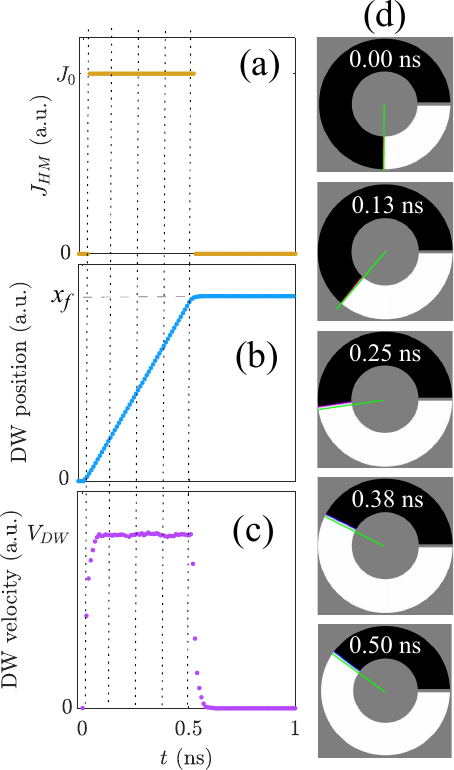}
\caption{Micromagnetic results of applying a uniform $J_{HM}=J_0=2$ TA/m$^2$ (a) showing the relative and final position $x_f$ (b) and velocity $V_{DW}$ (c), as a function of time $t$, of an UD DW in a curved strip ($w = 256$ nm and $r_e=384$ nm), at $T=260$ K. Insets (d) are snapshots of the magnetic configuration in sublattice $i=1$ at different times (highlighted by the vertical dotted lines in (a-c)). Green solid lines are for guiding the eye and are co-parallel with the strip radius.}  \label{teardrop_init}
\end{figure}

\subsection{Influence of width for a fixed curvature} 

\par In this study, the curvature is fixed ($r_e=384$ nm) and width ($w$) is varied from 56 nm to 296 nm in steps of 40 nm. Fig. 3 shows the results for the terminal DW velocity $(|V_{DW,i}|)$ as a function of the nominal density current $J_{HM}=j_{0}$, equivalent to the homogeneous density current in a straight strip with the same cross-section. In the next sections, we use the notation $J_{HM}=\text{J}$ for simplicity. Fig. 3 shows that temperature has a noticeable effect on the terminal velocity on the DW type equally, Up to Down domain (UD) or Down to Up domain (DU). As it was expected from previous work on straight FiM strips \cite{Alejos,PhysRevLett.121.057701_Liu_Siddiqui}, at $T_A$ the DW velocities are greater. Also, the DW velocities increase for narrower strips (red symbols). In fact, the observed trend is very similar to straight strips: the terminal velocity is maximum at the temperature of AMC, $T_{A}=260 $ K and significantly increased, exceeding 2000 m/s for the narrowest strip as compared to $\sim$1800 m/s for the widest. These results also suggest that the DWs velocities are equal for both types of DWs (UD and DU), which would lead to no distortion of the size of a domain between two adjacent DWs travelling along the curved strip. This result is significantly different from FM systems \cite{Alejos}.

\begin{figure}[t]
\centering 
\includegraphics[width=7.6cm]{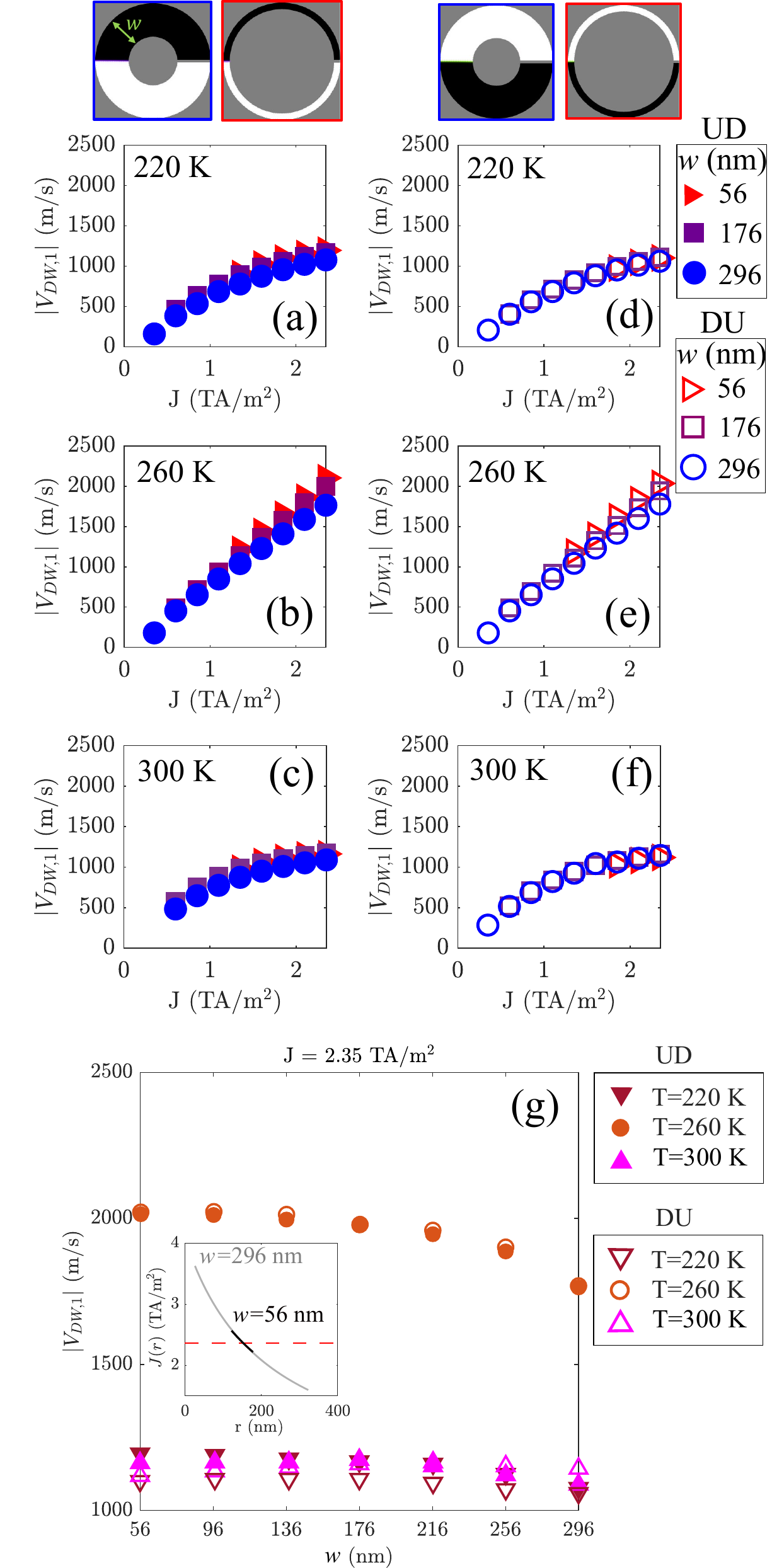}
\caption{Terminal velocities as a function of $\text{J}$ for a UD (a-c) and a DU (d-f) DW obtained for sub-lattice $i=1$ and for three different temperatures: below, above and at the AMC temperature (220 K, 300 K and 260 K, respectively). Strips for the two limiting cases are shown in the red and blue contour insets at the top. (g) Terminal velocities as a function of $w$ for a UD (full symbols) and a DU (open symbols) type wall for $\text{J}=2.35$ TA/m$^2$ and the three chosen temperatures. Inset in (g) shows $J(r)$ for two values of $w$. Red dashed line indicates $\text{J}=2.35$ TA/m$^2$.}  \label{teardrop_init}
\end{figure}

\par At $T\neq T_A$, the DW velocity is reduced either increasing or reducing temperature with respect to $T_A$, leading to velocities around 1100 m/s, generally regardless of the width and DW type. For a given $J$ value, as the strip gets wider, however, the velocity is slightly smaller but these differences are negligible (see Fig. 3(a), (c), (d), (f)). This result contrasts with that of a FM strip, where a greater difference of velocities between a DU and a UD along a curved strip was shown \cite{Alejos}.

\par To characterize the inertial motion of the DW, we evaluate the temporal evolution of the DW velocity ($V(t)$, computed from the spatial averaging of $m_{1,z}(t)$) as a function of time (or instant velocity) under a current square pulse of duration 0.1 ns and start at $t=0$.
The $\mu$m results can be fitted to the following exponentials: $V_{\infty}(1-e^{-t/\tau_r})$ during the duration of the pulse ($t\leq 0.1$ ns), and $V_{\infty}e^{-t/\tau_f}$, after the pulse ends ($t> 0.1$ ns), where $V_{\infty}$ is the DW terminal velocity (see Fig. 2(a)-(f)). The characteristic relaxation times $\tau_r$ (or rising time) and $\tau_f$ (or fall time) represent the duration of such transients and characterize the inertial motion of the DW. These parameters can be extracted by fitting the $\mu$m results to the exponentials (see solid curves in Fig. 4(a)).

\begin{figure}[t]
\centering 
\includegraphics[width=8.5cm]{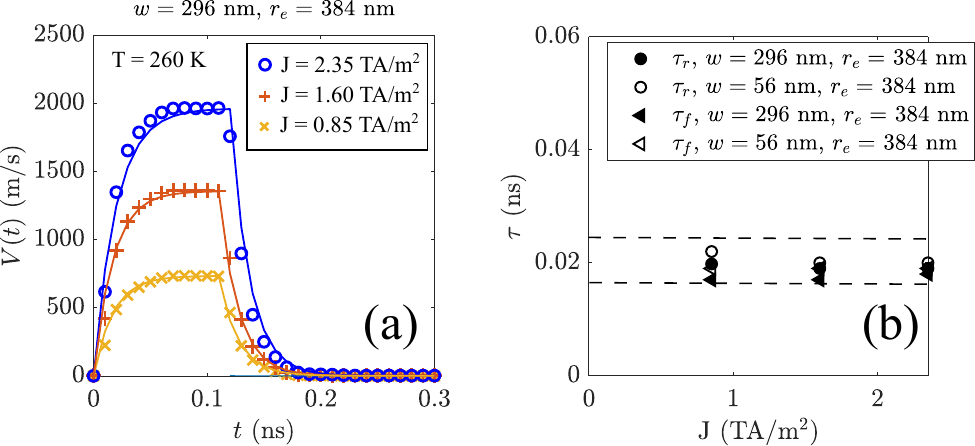}
\caption{Instantaneous DW velocities $V(t)$ for three values of $\text{J}$ and for a curved strip of $r_e = 384$ nm and $w=296$ nm at $T=T_{A}$ (a). Symbols are the $\mu$m data, from which $\tau$ (rise and fall times) are extracted for the narrowest and widest strips (b). Dashed lines in (b) are the upper and lower bounds of a 95$\%$ confidence interval.}  \label{teardrop_init}
\end{figure}

\par Although simulations were performed for both types of DWs, note that we only present here results for the DU type wall, for sake of simplicity. Identical results (not shown) were obtained for the UD DW. Fig. 4(a) show the \lq instantaneous' DW velocity $V(t)$ and the relaxation times $\tau$ for three selected values of $\text{J}$ (see solid symbols) at $T=T_A$ for the widest strip ($w=296$ nm). Solid lines are the exponential curves to which the obtained simulated data is fit. For each current, the minimal $\tau$ is expected for $T=T_A=260$ K.  Fig. 4(b) shows that $\tau_r$ and $\tau_f$ for the two limiting cases ($w=56$ nm and $w=296$ nm) are quantitatively similar, in the order of 0.02 ns, since they fall within the 95$\%$ confidence interval, set by the largest error bars obtained for $\tau$ from the fitted results, among all $\text{J}$. Also, all values are similar in order to the step-size used in simulations, 0.01 ns (see Fig. 4(d)).

\par Similar values of $\tau$ were obtained for strips of other widths. Relaxation times are not noticeably influenced by temperature, and they generally remain within the range of 0.01$\sim$0.03 ns for $T=200$ K and $T=300$ K. This is more than one order of magnitude smaller than in FM strips, the latter being about $\sim 1$ ns according to Ref.\cite{Thiaville}. Besides, the relaxation times of current-driven DWs in curved strips found here are in good agreement with those from field-driven or thermally driven DWs in antiferromagnetic straight strips, in the order of picoseconds \cite{PhysRevLett.117.017202,PhysRevLett.117.107201}.

\subsection{Influence of curvature for a fixed width} 

\par In this study, the strip width is fixed to $w=256$ nm and the curvature parameter $\rho$ is varied. In other words, the equivalent radii $r_{e}$ ($r_e=\rho ^{-1}$) is varied from 134 nm to 534 nm in steps of 50 nm. Fig. 5 shows the results for the terminal velocity $(|V_{DW,1}|)$ of DU and UD DWs, for several values of $r_e$ in nanometers, where the red (blue) curve corresponds to the smaller (greater) values, for three different temperatures.

\begin{figure}[t]
\centering 
\includegraphics[width=8cm]{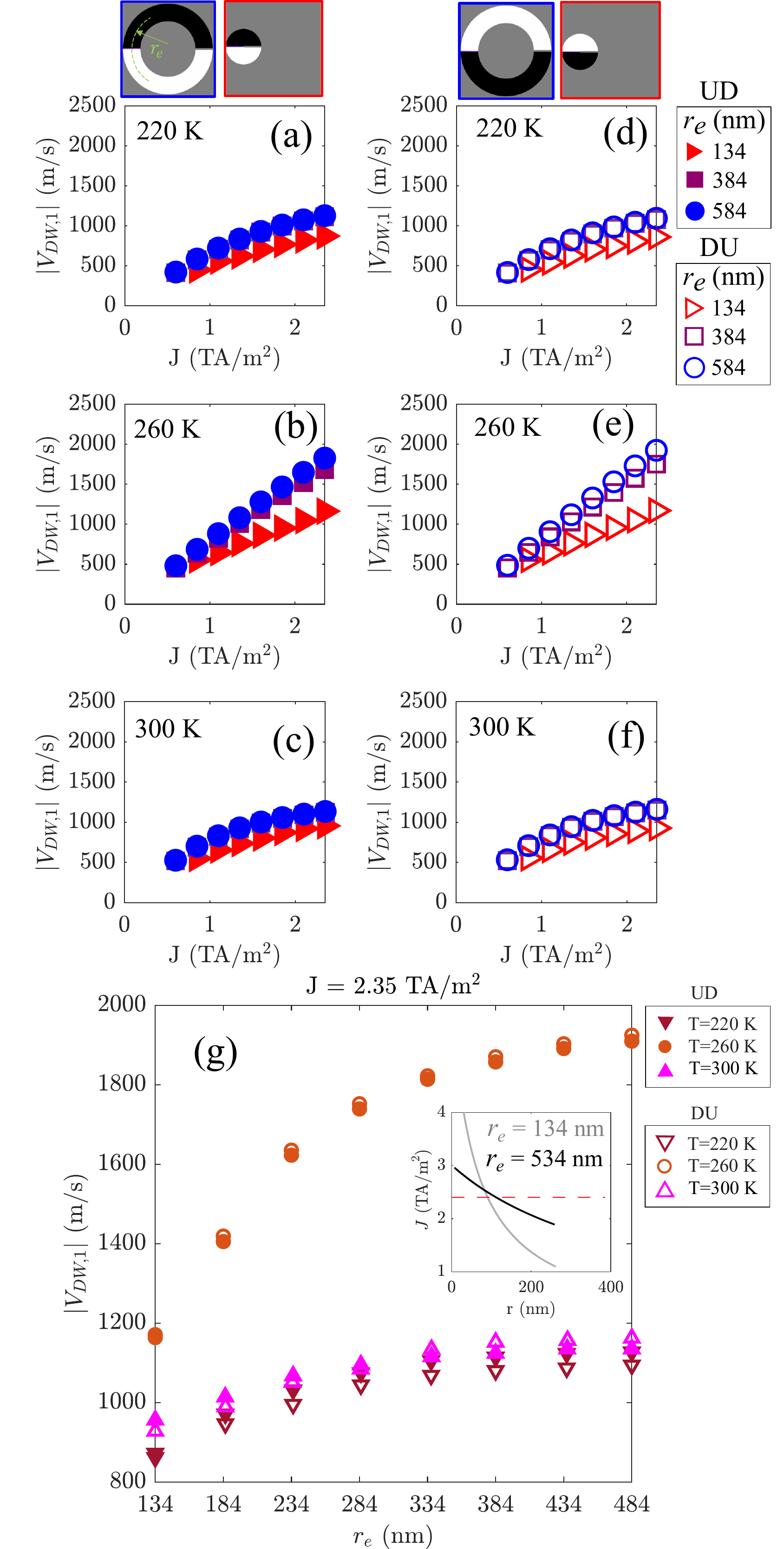}
\caption{Terminal velocities as a function of $\text{J}$ for a UD (a-c) and a DU (d-f) DW in the curved strip obtained for sub-lattice $i=1$ and for three different temperatures: below, above and at the AMC temperature (220 K, 300 K and 260 K, respectively). Strips for the two limiting cases are shown in the red and blue contour insets at the top. (g) Terminal velocities as a function of $r_e$ for a UD (full symbols) and a DU (open symbols) type wall for $\text{J}=2.35$ TA/m$^2$ and the three chosen temperatures. Inset in (g) shows $J(r)$ for two values of $r_e$. Red dashed line indicates $\text{J}=2.35$ TA/m$^2$.}  \label{teardrop_init}
\end{figure}

\par Fig. 5(a)-(f) shows that, at $T=T_{A}$ and for a given curvature, the DW velocities of DU and UD types are very similar for the whole range of currents explored. DW velocity reduces as the curvature increases (see Fig. 5(g)). It is worth noting that the latter cannot be a consequence of only a nonuniform $J(r)$ as defined in \cite{Alejos}. In fact, for curved-most strips ($r_e=134$ nm), the spatial-dependent density current $J(r)$ varies with $r$ similarly as it does for changing $w$ (see inset in Fig. 5(g) and in Fig. 3(g)), which would suggest similar variations to DW velocities as those found in Fig. 3(g). In other words, the impact of the non-uniform $J(r)$ is not so relevant to be the only source of the big differences between the DW velocities for large and small curvatures (orange symbols in Fig. 5(g) for $r_e=134$ nm and $r_e=484$ nm, respectively). Fig. 5(g) also shows that as the strip curvature is reduced, DW velocity converges to the straight strip case ($r_e \rightarrow \infty$). 

\par For $T \neq T_{A}$, the dependence of the DW velocity with temperature is minimal regardless of the DW type. As the strip curvature increases there is a prominent change in the maximal terminal DW velocity for both DW types. However, the relative difference of velocities is almost negligible. Therefore, results suggest that the strip curvature affects in a similar way to width, and equally, to both DWs. In other words, the terminal velocity is significantly reduced as curvature (or width) increases, while the differences between DWs remain negligible (see Fig. 5(g)). This behavior is even more pronounced at $T=T_A$. As discussed in section III.A, the latter would imply that the robustness of a transmitted bit, encoded in a domain between two DWs, can be optimised in such curved-most strips and reaches larger velocities in the strip.

\par Fig. 6(a) show the DW velocity as a function of time and for three selected values of $\text{J}$ for an effective radius of $r_e=534$ nm (least curved strip) and intermediate width $w=256$ nm, at $T=T_A$.  Results look quantitatively similar to those shown in Fig. 4(a), where $r_e$ was fixed to an intermediate value of $384$ nm.
\begin{figure}[t]
\centering 
\includegraphics[width=8.5cm]{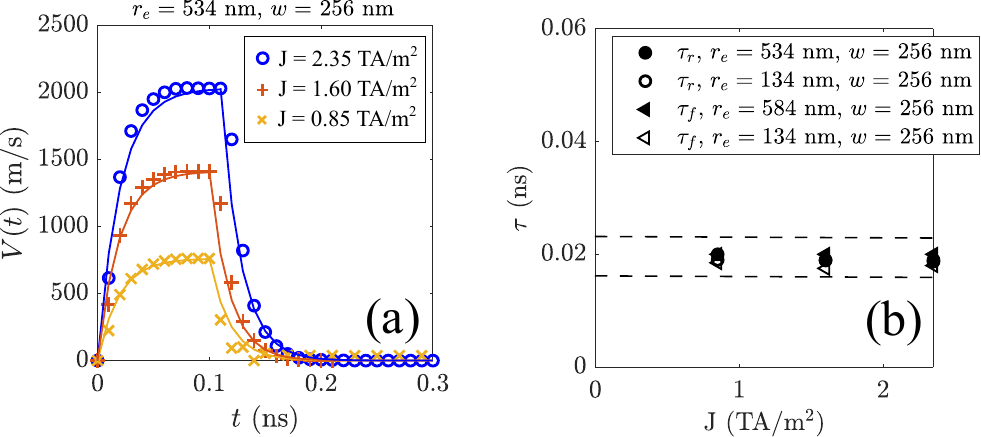}
\caption{Instantaneous DW velocities $V(t)$ for three values of $\text{J}$ and for a curved strip of $r_e = 584$ nm and $w=256$ nm at $T=T_{A}$ (a). Symbols are the $\mu$m data, from which $\tau$ (rise and fall times) are extracted for the narrowest and widest strips (b). Dashed lines in (b) are the upper and lower bounds of a 95$\%$ confidence interval.}  \label{teardrop_init}
\end{figure}
As in Fig. 4(b), Fig. 6(b) shows that $\tau_r$ and $\tau_f$ for the two limiting cases ($r_e=134$ nm and $r_e=534$ nm) are quantitatively similar, in the order of 0.02 ns. For all the FiM curved strips explored at, above and below AMC, $\tau_r$ and $\tau_f$ remain within the range of 0.01$\sim$0.03 ns, approximately one order of magnitude less than their FM counterparts. This is in good agreement with results presented in the previous section and other work in straight strips \cite{inertia}, which further supports the negligible inertia of DWs in such FiM systems.

\subsection{Discussion on the effective influence of a curved shape on the wall velocity} 

\par A non-uniform current distribution is expected to influence the terminal velocity of the DW for a given curvature, specially for wide curved strips \cite{Alejos}. In this section, to explore further the degree of influence of the non-uniform current, equivalent studies on $w$ and $\rho$ on a straight strip with an artificially implemented non-uniform $\text{J}(r=y)$ at $T=T_A$ are performed. A straight strip is a bounding case for a curved strip that shows no effective curvature ($r_e \rightarrow \infty, \rho \rightarrow 0$)  and an homogeneous density current $J = J_0$. Therefore, we explore whether \lq curvature ($\rho$) effects' are mainly dominated by the intrinsic inhomogenous current, or whether they can also arise from the curved shape itself \cite{yang_2017}. We aim to discern the actual influence of an inhomogeneous current, as part of an more global effect due to the curved shape.  For the following study, and since the straight shape must be retained, $r_e$ (or equivalently $\rho$) is artificially modified in the non-uniform current distribution expression: $\vec{\text{J}}(y,w,r_e)$ \cite{Alejos} in the $x$-direction, as if the strip was curved. Note that $\rho$ represents the inverse of the averaged or effective curvature radius of the strip ($\rho=r_e^{-1}$) and not the cylindrical radial coordinate ($r$) in the system.

\begin{figure}[t]
\centering 
\includegraphics[width=8.5cm]{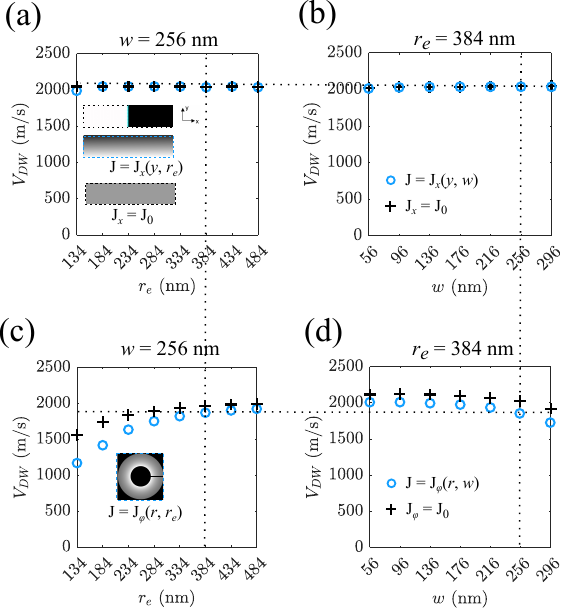}
\caption{DW terminal velocities in a straight strip for an inhomogeneous $J(y,w,r_e)$ (blue circles) and for a uniform $J_0$ (black crosses) as a function of radius $r_e$ (a) and width $w$ (b) at same temperature ($T=260$ K). Insets in (a) show the magnetic configuration of sublattice $i=1$ at $t=0$ and schematics of the current spatial distribution in the strips as examples. (b) DW terminal velocities in a curved strips with the same parameters as a function of radius $r_e$ (c) and width $w$ (d). As an example, inset in (c) shows the radial dependence of an inhomogenous current distribution in such a strip. Dotted lines highlight the cases where the two geometrical parameters ($w$ and $r_e$) are coincident among all the studies.}  \label{teardrop_init}
\end{figure}

\par Fig. 7(a-b) show the velocity of an UD wall in the straight strip for $J = J_0=2.35$ TA/m$^2$ (black crosses) and for an inhomogeneous $J(y)$ (blue circles), varying $r_e$ in $J(y,r_e)$ (see insets) for a fixed width $w=256$ nm, and varying width ($w$) for a fixed $r_e$, $J(y,r_e=384)$. While it is expected that an inhomogeneous $J(y,w,r_e)$ will influence the DW velocity (especially for curved-most strips, see $r_e=134$ nm in (a)), the differences with the case of $J=J_0$ are almost negligible. Fig. 7(c-d) show results for curved strips. For these cases, $w$ and $r_e$ are naturally varied in $J(r,w,r_e)$ by modifying the shape itself. From the standpoint of the applied current, this is expected to be equivalent to doing it by changing the shape itself. Results from an inhomogeneous current (blue circles, reproduced from Fig. 3(g) and 5(g)) consistently tend to converge to the straight strip as $\rho$ is reduced. The strip is straight when $\rho=0$  ($r_e \rightarrow \infty$).

\par When the strip is either straight (a-b) or curved (c-d), for both studies (fixing $w$ and varying $r_e$ or vice-versa), the DW velocity is found to be the same when the geometrical parameters are coincident, i.e., $w=256$ nm and $r_e=384$ nm, as expected (see horizontal dotted lines in (a-b) or (c-d)). However, when the shape is different, even in the cases when $w$ and $r_e$ are coincident (and therefore, $J(w,r_e)$ is expected to also be the same), different DW velocities are obtained (see vertical dotted lines in (a-c) and (b-d)) below and slightly above 2000 m/s, respectively. Moreover, by modifying either $r_e$ or $w$ in the curved strip by directly changing its shape (see (c-d) and previous sections), there is a clear larger impact on DW velocity, than by artificially (but equivalently) modifying $r_e$ or $w$ in the straight strip (see (a-c)). In the curved strips, the trend when varying $r_e$ or $w$ is qualitatively similar regardless to the homogeneity of the applied $J$ (see black crosses and blue circles in either (c) or (d)). Since $\text{J}(r)$ is modeled in an equivalent way in all studies by simply changing $\text{J}(r)=\text{J}(y)$ for the straight strip, marked differences between the wall velocity in (a-b) (straight strip) and (c-d) (curved strip), specially for very curved strips, suggest that not only the non-uniform $\text{J}(r)$ is influencing the UD DW motion. 

\par Our results suggest that the curved shape may have an intrinsic influence on the wall velocity, manifested as a more marked dependence with $w$ and $r_e$, regardless of the inhomogeneity of the current density (see Fig.(c-d)), as the shape becomes more curved. An influence due to the inhomogeneous current, still appears naturally in the curved strip, but may have a lesser impact compared to other geometrical factors (see differences between black crosses and blue circles).

\section{Discussion and summary}
\par We have provided a study on DW motion in curved FiM strips, particularised to one of the two strongly coupled sub-lattices, for  three different temperatures, and as a function of geometrical parameters for a HM/FiM/Ox multilayer structure. We observe an absence of tilting of the DW and domain distortion at different temperatures, 40 K above and below the angular momentum compensation temperature.

\par Width and curvature effects on the DW velocity are discussed. Besides contributions from a non-uniform $\text{J}(r)$, there is an overall significant influence from the shape of the strip itself on DW velocity. This implies that, for a fixed temperature (or composition), DW velocity can be optimised by optimising the geometrical parameters of the curved strip. The relative differences between a DU and a UD walls are marginal in general. In other words, geometrical factors affect them almost equally, which is positive for a robust transmission of a bit encoded in an Up or Down domain between two adjacent DWs. With reducing current, differences in velocities between curved and straight strips are still minimised at the expense of slower DWs. Similar effect is observed as width or curvature is increased. This is beneficial for designing intricate 2D circuit tracks combining curved and straight sections, while preserving DW velocities still larger than those found in their FM counterparts. Also, DWs in a curved FiM strip show a negligible inertia in contrast to their FM counterparts ($\tau_{FiM}<<\tau_{FM}$) for all the explored scenarios. The DWs start to move and stop almost immediately ($\tau_{FiM} \sim 0.02$ ns) after the application or removal of current.

\par Considering the obtained results altogether and assuming $T\neq T_{A}$, which will be most of the experimental cases at room temperature ($T=300$ K), our study allows us to conclude that narrow enough FiM strips are ideal candidates for designing curved tracks for 2D spintronic circuits of an arbitrary shape based on CDDWM, where bits are encoded in domains separated by walls. This is due to very fast rise and fall times ($\tau_r \sim \tau_f <<0.1$ ns), high velocities ($V_{DW}> 1000 $ m/s) and negligible distortion of the two types of DWs (UD and DU) in all the scenarios explored in this work. Greater DW terminal velocities and smaller time responses in curved FiM strips than those in their FM counterparts are obtained. These results can help in the further research, development and improvement of FiM-based spintronic circuitry that may require compactness and high-speed functionality with high robustness to DW (and/or domain) distortion.
 
\section{Acknowledgements}
\par This work was supported by project SA114P20 from Junta de Castilla y Leon (JCyL), and partially supported by projects SA299P18 from JCyL, MAT2017-87072-C4-1-P and PID2020-117024GB-C41 from the Ministry of Economy, Spanish government, and MAGNEFI, from the European Commission (European Union). All data created during this research are openly available from the University of Salamanca's institutional repository at https://gredos.usal.es/handle/10366/138189

\bibliography{library}

\begin{thebibliography}{24}%
\makeatletter
\providecommand \@ifxundefined [1]{%
 \@ifx{#1\undefined}
}%
\providecommand \@ifnum [1]{%
 \ifnum #1\expandafter \@firstoftwo
 \else \expandafter \@secondoftwo
 \fi
}%
\providecommand \@ifx [1]{%
 \ifx #1\expandafter \@firstoftwo
 \else \expandafter \@secondoftwo
 \fi
}%
\providecommand \natexlab [1]{#1}%
\providecommand \enquote  [1]{``#1''}%
\providecommand \bibnamefont  [1]{#1}%
\providecommand \bibfnamefont [1]{#1}%
\providecommand \citenamefont [1]{#1}%
\providecommand \href@noop [0]{\@secondoftwo}%
\providecommand \href [0]{\begingroup \@sanitize@url \@href}%
\providecommand \@href[1]{\@@startlink{#1}\@@href}%
\providecommand \@@href[1]{\endgroup#1\@@endlink}%
\providecommand \@sanitize@url [0]{\catcode `\\12\catcode `\$12\catcode
  `\&12\catcode `\#12\catcode `\^12\catcode `\_12\catcode `\%12\relax}%
\providecommand \@@startlink[1]{}%
\providecommand \@@endlink[0]{}%
\providecommand \url  [0]{\begingroup\@sanitize@url \@url }%
\providecommand \@url [1]{\endgroup\@href {#1}{\urlprefix }}%
\providecommand \urlprefix  [0]{URL }%
\providecommand \Eprint [0]{\href }%
\providecommand \doibase [0]{http://dx.doi.org/}%
\providecommand \selectlanguage [0]{\@gobble}%
\providecommand \bibinfo  [0]{\@secondoftwo}%
\providecommand \bibfield  [0]{\@secondoftwo}%
\providecommand \translation [1]{[#1]}%
\providecommand \BibitemOpen [0]{}%
\providecommand \bibitemStop [0]{}%
\providecommand \bibitemNoStop [0]{.\EOS\space}%
\providecommand \EOS [0]{\spacefactor3000\relax}%
\providecommand \BibitemShut  [1]{\csname bibitem#1\endcsname}%
\let\auto@bib@innerbib\@empty
\bibitem [{\citenamefont {Hubert}\ and\ \citenamefont
  {Schäfer}(1998)}]{hubert98}%
  \BibitemOpen
  \bibfield  {author} {\bibinfo {author} {\bibfnamefont {A.}~\bibnamefont
  {Hubert}}\ and\ \bibinfo {author} {\bibfnamefont {R.}~\bibnamefont
  {Schäfer}},\ }in\ \href {\doibase 10.1007/978-3-540-85054-0} {\emph
  {\bibinfo {booktitle} {Magnetic Domains: The Analysis of Magnetic
  Microstructures}}}\ (\bibinfo  {publisher} {Springer-Verlag Berlin
  Heidelberg},\ \bibinfo {year} {1998})\BibitemShut {NoStop}%
\bibitem [{\citenamefont {Parkin}(2004)}]{Parkinpat}%
  \BibitemOpen
  \bibfield  {author} {\bibinfo {author} {\bibfnamefont {S.~S.~P.}\
  \bibnamefont {Parkin}},\ }\href@noop {} {\enquote {\bibinfo {title}
  {Shiftable magnetic shift register and method of using the same},}\ }
  (\bibinfo {year} {US6834005B1, 2004})\BibitemShut {NoStop}%
\bibitem [{\citenamefont {Parkin}\ and\ \citenamefont
  {Yang}(2015)}]{parkinracetrack}%
  \BibitemOpen
  \bibfield  {author} {\bibinfo {author} {\bibfnamefont {S.}~\bibnamefont
  {Parkin}}\ and\ \bibinfo {author} {\bibfnamefont {S.-H.}\ \bibnamefont
  {Yang}},\ }\href {\doibase 10.1038/nnano.2015.41} {\bibfield  {journal}
  {\bibinfo  {journal} {Nature nanotechnology}\ }\textbf {\bibinfo {volume}
  {10}},\ \bibinfo {pages} {195—198} (\bibinfo {year} {2015})}\BibitemShut
  {NoStop}%
\bibitem [{\citenamefont {Miron}\ \emph {et~al.}(2011)\citenamefont {Miron},
  \citenamefont {Moore}, \citenamefont {Szambolics}, \citenamefont
  {Buda-Prejbeanu}, \citenamefont {Auffret}, \citenamefont {Rodmacq},
  \citenamefont {Pizzini}, \citenamefont {Vogel}, \citenamefont {Bonfim},
  \citenamefont {Schuhl},\ and\ \citenamefont {Gaudin}}]{miron}%
  \BibitemOpen
  \bibfield  {author} {\bibinfo {author} {\bibfnamefont {I.}~\bibnamefont
  {Miron}}, \bibinfo {author} {\bibfnamefont {T.}~\bibnamefont {Moore}},
  \bibinfo {author} {\bibfnamefont {H.}~\bibnamefont {Szambolics}}, \bibinfo
  {author} {\bibfnamefont {L.}~\bibnamefont {Buda-Prejbeanu}}, \bibinfo
  {author} {\bibfnamefont {S.}~\bibnamefont {Auffret}}, \bibinfo {author}
  {\bibfnamefont {B.}~\bibnamefont {Rodmacq}}, \bibinfo {author} {\bibfnamefont
  {S.}~\bibnamefont {Pizzini}}, \bibinfo {author} {\bibfnamefont
  {J.}~\bibnamefont {Vogel}}, \bibinfo {author} {\bibfnamefont
  {M.}~\bibnamefont {Bonfim}}, \bibinfo {author} {\bibfnamefont
  {A.}~\bibnamefont {Schuhl}}, \ and\ \bibinfo {author} {\bibfnamefont
  {G.}~\bibnamefont {Gaudin}},\ }\href {\doibase 10.1038/nmat3020} {\bibfield
  {journal} {\bibinfo  {journal} {Nature materials}\ }\textbf {\bibinfo
  {volume} {10}},\ \bibinfo {pages} {419} (\bibinfo {year} {2011})}\BibitemShut
  {NoStop}%
\bibitem [{\citenamefont {Emori}\ \emph {et~al.}(2013)\citenamefont {Emori},
  \citenamefont {Bauer}, \citenamefont {Ahn}, \citenamefont {Martinez},\ and\
  \citenamefont {Beach}}]{emori}%
  \BibitemOpen
  \bibfield  {author} {\bibinfo {author} {\bibfnamefont {S.}~\bibnamefont
  {Emori}}, \bibinfo {author} {\bibfnamefont {U.}~\bibnamefont {Bauer}},
  \bibinfo {author} {\bibfnamefont {S.-M.}\ \bibnamefont {Ahn}}, \bibinfo
  {author} {\bibfnamefont {E.}~\bibnamefont {Martinez}}, \ and\ \bibinfo
  {author} {\bibfnamefont {G.~S.~D.}\ \bibnamefont {Beach}},\ }\href {\doibase
  10.1038/nmat3675} {\bibfield  {journal} {\bibinfo  {journal} {Nature
  materials}\ }\textbf {\bibinfo {volume} {12}},\ \bibinfo {pages} {611—616}
  (\bibinfo {year} {2013})}\BibitemShut {NoStop}%
\bibitem [{\citenamefont {Haazen}\ \emph {et~al.}(2013)\citenamefont {Haazen},
  \citenamefont {Mur{\'e}}, \citenamefont {Franken}, \citenamefont {Lavrijsen},
  \citenamefont {Swagten},\ and\ \citenamefont {Koopmans}}]{haazen}%
  \BibitemOpen
  \bibfield  {author} {\bibinfo {author} {\bibfnamefont {P.}~\bibnamefont
  {Haazen}}, \bibinfo {author} {\bibfnamefont {E.}~\bibnamefont {Mur{\'e}}},
  \bibinfo {author} {\bibfnamefont {J.}~\bibnamefont {Franken}}, \bibinfo
  {author} {\bibfnamefont {R.}~\bibnamefont {Lavrijsen}}, \bibinfo {author}
  {\bibfnamefont {H.}~\bibnamefont {Swagten}}, \ and\ \bibinfo {author}
  {\bibfnamefont {B.}~\bibnamefont {Koopmans}},\ }\href {\doibase
  10.1038/NMAT3553} {\bibfield  {journal} {\bibinfo  {journal} {Nature
  Materials}\ }\textbf {\bibinfo {volume} {12}},\ \bibinfo {pages} {299}
  (\bibinfo {year} {2013})}\BibitemShut {NoStop}%
\bibitem [{\citenamefont {Ryu}\ \emph {et~al.}(2013)\citenamefont {Ryu},
  \citenamefont {Thomas}, \citenamefont {Yang},\ and\ \citenamefont
  {Parkin}}]{ryu}%
  \BibitemOpen
  \bibfield  {author} {\bibinfo {author} {\bibfnamefont {K.-S.}\ \bibnamefont
  {Ryu}}, \bibinfo {author} {\bibfnamefont {L.}~\bibnamefont {Thomas}},
  \bibinfo {author} {\bibfnamefont {S.-H.}\ \bibnamefont {Yang}}, \ and\
  \bibinfo {author} {\bibfnamefont {S.}~\bibnamefont {Parkin}},\ }\href
  {\doibase 10.1038/nnano.2013.102} {\bibfield  {journal} {\bibinfo  {journal}
  {Nature nanotechnology}\ }\textbf {\bibinfo {volume} {8}} (\bibinfo {year}
  {2013}),\ 10.1038/nnano.2013.102}\BibitemShut {NoStop}%
\bibitem [{\citenamefont {Torrejon}\ \emph {et~al.}(2016)\citenamefont
  {Torrejon}, \citenamefont {Martinez},\ and\ \citenamefont
  {Hayashi}}]{inertia}%
  \BibitemOpen
  \bibfield  {author} {\bibinfo {author} {\bibfnamefont {J.}~\bibnamefont
  {Torrejon}}, \bibinfo {author} {\bibfnamefont {E.}~\bibnamefont {Martinez}},
  \ and\ \bibinfo {author} {\bibfnamefont {M.}~\bibnamefont {Hayashi}},\ }\href
  {\doibase 10.1038/ncomms13533} {\bibfield  {journal} {\bibinfo  {journal}
  {Nature communications}\ }\textbf {\bibinfo {volume} {7}},\ \bibinfo {pages}
  {13533} (\bibinfo {year} {2016})}\BibitemShut {NoStop}%
\bibitem [{\citenamefont {Dzyaloshinsky}(1958)}]{DZYALOSHINSKY1958241}%
  \BibitemOpen
  \bibfield  {author} {\bibinfo {author} {\bibfnamefont {I.}~\bibnamefont
  {Dzyaloshinsky}},\ }\href {\doibase
  https://doi.org/10.1016/0022-3697(58)90076-3} {\bibfield  {journal} {\bibinfo
   {journal} {Journal of Physics and Chemistry of Solids}\ }\textbf {\bibinfo
  {volume} {4}},\ \bibinfo {pages} {241} (\bibinfo {year} {1958})}\BibitemShut
  {NoStop}%
\bibitem [{\citenamefont {Garg}\ \emph {et~al.}(2017)\citenamefont {Garg},
  \citenamefont {Yang}, \citenamefont {Phung}, \citenamefont {Pushp},\ and\
  \citenamefont {Parkin}}]{garg}%
  \BibitemOpen
  \bibfield  {author} {\bibinfo {author} {\bibfnamefont {C.}~\bibnamefont
  {Garg}}, \bibinfo {author} {\bibfnamefont {S.-H.}\ \bibnamefont {Yang}},
  \bibinfo {author} {\bibfnamefont {T.}~\bibnamefont {Phung}}, \bibinfo
  {author} {\bibfnamefont {A.}~\bibnamefont {Pushp}}, \ and\ \bibinfo {author}
  {\bibfnamefont {S.}~\bibnamefont {Parkin}},\ }\href {\doibase
  10.1126/sciadv.1602804} {\bibfield  {journal} {\bibinfo  {journal} {Science
  Advances}\ }\textbf {\bibinfo {volume} {3}},\ \bibinfo {pages} {e1602804}
  (\bibinfo {year} {2017})}\BibitemShut {NoStop}%
\bibitem [{\citenamefont {Alejos}\ \emph {et~al.}(2020)\citenamefont {Alejos},
  \citenamefont {Raposo},\ and\ \citenamefont {Martínez}}]{Alejos}%
  \BibitemOpen
  \bibfield  {author} {\bibinfo {author} {\bibfnamefont {O.}~\bibnamefont
  {Alejos}}, \bibinfo {author} {\bibfnamefont {V.}~\bibnamefont {Raposo}}, \
  and\ \bibinfo {author} {\bibfnamefont {E.}~\bibnamefont {Martínez}},\
  }\enquote {\bibinfo {title} {Domain wall motion in magnetic nanostrips},}\
  in\ \href {\doibase https://doi.org/10.1002/9783527603978.mst0459} {\emph
  {\bibinfo {booktitle} {Materials Science and Technology}}}\ (\bibinfo
  {publisher} {American Cancer Society},\ \bibinfo {year} {2020})\ pp.\
  \bibinfo {pages} {1--49}\BibitemShut {NoStop}%
\bibitem [{\citenamefont {Yang}\ \emph {et~al.}(2015)\citenamefont {Yang},
  \citenamefont {Ryu},\ and\ \citenamefont {Parkin}}]{yang}%
  \BibitemOpen
  \bibfield  {author} {\bibinfo {author} {\bibfnamefont {S.-H.}\ \bibnamefont
  {Yang}}, \bibinfo {author} {\bibfnamefont {K.-S.}\ \bibnamefont {Ryu}}, \
  and\ \bibinfo {author} {\bibfnamefont {S.}~\bibnamefont {Parkin}},\ }\href
  {\doibase 10.1038/nnano.2014.324} {\bibfield  {journal} {\bibinfo  {journal}
  {Nature nanotechnology}\ }\textbf {\bibinfo {volume} {10}} (\bibinfo {year}
  {2015}),\ 10.1038/nnano.2014.324}\BibitemShut {NoStop}%
\bibitem [{\citenamefont {Siddiqui}\ \emph {et~al.}(2018)\citenamefont
  {Siddiqui}, \citenamefont {Han}, \citenamefont {Finley}, \citenamefont
  {Ross},\ and\ \citenamefont {Liu}}]{PhysRevLett.121.057701_Liu_Siddiqui}%
  \BibitemOpen
  \bibfield  {author} {\bibinfo {author} {\bibfnamefont {S.~A.}\ \bibnamefont
  {Siddiqui}}, \bibinfo {author} {\bibfnamefont {J.}~\bibnamefont {Han}},
  \bibinfo {author} {\bibfnamefont {J.~T.}\ \bibnamefont {Finley}}, \bibinfo
  {author} {\bibfnamefont {C.~A.}\ \bibnamefont {Ross}}, \ and\ \bibinfo
  {author} {\bibfnamefont {L.}~\bibnamefont {Liu}},\ }\href {\doibase
  10.1103/PhysRevLett.121.057701} {\bibfield  {journal} {\bibinfo  {journal}
  {Phys. Rev. Lett.}\ }\textbf {\bibinfo {volume} {121}},\ \bibinfo {pages}
  {057701} (\bibinfo {year} {2018})}\BibitemShut {NoStop}%
\bibitem [{\citenamefont {Caretta}\ \emph {et~al.}(2018)\citenamefont
  {Caretta}, \citenamefont {Mann}, \citenamefont {B{\"u}ttner}, \citenamefont
  {Ueda}, \citenamefont {Pfau}, \citenamefont {G{\"u}nther}, \citenamefont
  {Hessing}, \citenamefont {Churikova}, \citenamefont {Klose}, \citenamefont
  {Schneider}, \citenamefont {Engel}, \citenamefont {Marcus}, \citenamefont
  {Bono}, \citenamefont {Bagschik}, \citenamefont {Eisebitt},\ and\
  \citenamefont {Beach}}]{Caretta2018FastCD}%
  \BibitemOpen
  \bibfield  {author} {\bibinfo {author} {\bibfnamefont {L.}~\bibnamefont
  {Caretta}}, \bibinfo {author} {\bibfnamefont {M.}~\bibnamefont {Mann}},
  \bibinfo {author} {\bibfnamefont {F.}~\bibnamefont {B{\"u}ttner}}, \bibinfo
  {author} {\bibfnamefont {K.}~\bibnamefont {Ueda}}, \bibinfo {author}
  {\bibfnamefont {B.}~\bibnamefont {Pfau}}, \bibinfo {author} {\bibfnamefont
  {C.}~\bibnamefont {G{\"u}nther}}, \bibinfo {author} {\bibfnamefont
  {P.}~\bibnamefont {Hessing}}, \bibinfo {author} {\bibfnamefont
  {A.}~\bibnamefont {Churikova}}, \bibinfo {author} {\bibfnamefont
  {C.}~\bibnamefont {Klose}}, \bibinfo {author} {\bibfnamefont
  {M.}~\bibnamefont {Schneider}}, \bibinfo {author} {\bibfnamefont
  {D.}~\bibnamefont {Engel}}, \bibinfo {author} {\bibfnamefont
  {C.}~\bibnamefont {Marcus}}, \bibinfo {author} {\bibfnamefont
  {D.}~\bibnamefont {Bono}}, \bibinfo {author} {\bibfnamefont {K.}~\bibnamefont
  {Bagschik}}, \bibinfo {author} {\bibfnamefont {S.}~\bibnamefont {Eisebitt}},
  \ and\ \bibinfo {author} {\bibfnamefont {G.}~\bibnamefont {Beach}},\
  }\href@noop {} {\bibfield  {journal} {\bibinfo  {journal} {Nature
  Nanotechnology}\ }\textbf {\bibinfo {volume} {13}},\ \bibinfo {pages} {1154}
  (\bibinfo {year} {2018})}\BibitemShut {NoStop}%
\bibitem [{\citenamefont {Martínez}\ \emph {et~al.}(2019)\citenamefont
  {Martínez}, \citenamefont {Raposo},\ and\ \citenamefont {Óscar
  Alejos}}]{MARTINEZ2019165545}%
  \BibitemOpen
  \bibfield  {author} {\bibinfo {author} {\bibfnamefont {E.}~\bibnamefont
  {Martínez}}, \bibinfo {author} {\bibfnamefont {V.}~\bibnamefont {Raposo}}, \
  and\ \bibinfo {author} {\bibnamefont {Óscar Alejos}},\ }\href {\doibase
  https://doi.org/10.1016/j.jmmm.2019.165545} {\bibfield  {journal} {\bibinfo
  {journal} {Journal of Magnetism and Magnetic Materials}\ }\textbf {\bibinfo
  {volume} {491}},\ \bibinfo {pages} {165545} (\bibinfo {year}
  {2019})}\BibitemShut {NoStop}%
\bibitem [{\citenamefont {{S. Arpaci, V. Lopez-Dominguez, J. Shi, L.
  S{\'a}nchez-Tejerina, F. Garesci, C. Wang, X. Yan, V. K. Sangwan, M. A.
  Grayson, M. C. Hersam, G. Finocchio and P. Khalili
  Amiri}}(2021)}]{arpaci2021}%
  \BibitemOpen
  \bibfield  {author} {\bibinfo {author} {\bibnamefont {{S. Arpaci, V.
  Lopez-Dominguez, J. Shi, L. S{\'a}nchez-Tejerina, F. Garesci, C. Wang, X.
  Yan, V. K. Sangwan, M. A. Grayson, M. C. Hersam, G. Finocchio and P. Khalili
  Amiri}}},\ }\href {\doibase 10.1038/s41467-021-24237-y} {\bibfield  {journal}
  {\bibinfo  {journal} {Nature Communications}\ }\textbf {\bibinfo {volume}
  {12}} (\bibinfo {year} {2021}),\ 10.1038/s41467-021-24237-y}\BibitemShut
  {NoStop}%
\bibitem [{\citenamefont {Landau}\ and\ \citenamefont
  {Lifshitz}(1935)}]{LANDAU199251}%
  \BibitemOpen
  \bibfield  {author} {\bibinfo {author} {\bibfnamefont {L.}~\bibnamefont
  {Landau}}\ and\ \bibinfo {author} {\bibfnamefont {E.}~\bibnamefont
  {Lifshitz}},\ }\href@noop {} {\bibfield  {journal} {\bibinfo  {journal}
  {Phys. Zeitsch. der Sow.}\ }\textbf {\bibinfo {volume} {8}},\ \bibinfo
  {pages} {153–169} (\bibinfo {year} {1935})}\BibitemShut {NoStop}%
\bibitem [{\citenamefont {Martinez}\ \emph {et~al.}(2014)\citenamefont
  {Martinez}, \citenamefont {Perez}, \citenamefont {Torres}, \citenamefont
  {Emori},\ and\ \citenamefont {Beach}}]{martinez2014}%
  \BibitemOpen
  \bibfield  {author} {\bibinfo {author} {\bibfnamefont {E.}~\bibnamefont
  {Martinez}}, \bibinfo {author} {\bibfnamefont {N.}~\bibnamefont {Perez}},
  \bibinfo {author} {\bibfnamefont {L.}~\bibnamefont {Torres}}, \bibinfo
  {author} {\bibfnamefont {S.}~\bibnamefont {Emori}}, \ and\ \bibinfo {author}
  {\bibfnamefont {G.~S.~D.}\ \bibnamefont {Beach}},\ }\href {\doibase
  10.1063/1.4881778} {\bibfield  {journal} {\bibinfo  {journal} {Journal of
  Applied Physics}\ }\textbf {\bibinfo {volume} {115}} (\bibinfo {year}
  {2014}),\ 10.1063/1.4881778}\BibitemShut {NoStop}%
\bibitem [{\citenamefont {Slonczewski}(1996)}]{slonczewski1996current}%
  \BibitemOpen
  \bibfield  {author} {\bibinfo {author} {\bibfnamefont {J.~C.}\ \bibnamefont
  {Slonczewski}},\ }\href@noop {} {\bibfield  {journal} {\bibinfo  {journal}
  {Journal of Magnetism and Magnetic Materials}\ }\textbf {\bibinfo {volume}
  {159}},\ \bibinfo {pages} {L1} (\bibinfo {year} {1996})}\BibitemShut
  {NoStop}%
\bibitem [{\citenamefont {Alejos}\ \emph {et~al.}(2018)\citenamefont {Alejos},
  \citenamefont {Raposo}, \citenamefont {Sanchez-Tejerina}, \citenamefont
  {Tomasello}, \citenamefont {Finocchio},\ and\ \citenamefont
  {Martinez}}]{model2018}%
  \BibitemOpen
  \bibfield  {author} {\bibinfo {author} {\bibfnamefont {O.}~\bibnamefont
  {Alejos}}, \bibinfo {author} {\bibfnamefont {V.}~\bibnamefont {Raposo}},
  \bibinfo {author} {\bibfnamefont {L.}~\bibnamefont {Sanchez-Tejerina}},
  \bibinfo {author} {\bibfnamefont {R.}~\bibnamefont {Tomasello}}, \bibinfo
  {author} {\bibfnamefont {G.}~\bibnamefont {Finocchio}}, \ and\ \bibinfo
  {author} {\bibfnamefont {E.}~\bibnamefont {Martinez}},\ }\href
  {www.scopus.com} {\bibfield  {journal} {\bibinfo  {journal} {Journal of
  Applied Physics}\ }\textbf {\bibinfo {volume} {123}} (\bibinfo {year}
  {2018})}\BibitemShut {NoStop}%
\bibitem [{\citenamefont {{Thiaville}}\ \emph {et~al.}(2007)\citenamefont
  {{Thiaville}}, \citenamefont {{Nakatani}}, \citenamefont {{Pi{\'e}chon}},
  \citenamefont {{Miltat}},\ and\ \citenamefont {{Ono}}}]{Thiaville}%
  \BibitemOpen
  \bibfield  {author} {\bibinfo {author} {\bibfnamefont {A.}~\bibnamefont
  {{Thiaville}}}, \bibinfo {author} {\bibfnamefont {Y.}~\bibnamefont
  {{Nakatani}}}, \bibinfo {author} {\bibfnamefont {F.}~\bibnamefont
  {{Pi{\'e}chon}}}, \bibinfo {author} {\bibfnamefont {J.}~\bibnamefont
  {{Miltat}}}, \ and\ \bibinfo {author} {\bibfnamefont {T.}~\bibnamefont
  {{Ono}}},\ }\href {\doibase 10.1140/epjb/e2007-00320-3} {\bibfield  {journal}
  {\bibinfo  {journal} {European Physical Journal B}\ }\textbf {\bibinfo
  {volume} {60}},\ \bibinfo {pages} {15} (\bibinfo {year} {2007})}\BibitemShut
  {NoStop}%
\bibitem [{\citenamefont {Gomonay}\ \emph {et~al.}(2016)\citenamefont
  {Gomonay}, \citenamefont {Jungwirth},\ and\ \citenamefont
  {Sinova}}]{PhysRevLett.117.017202}%
  \BibitemOpen
  \bibfield  {author} {\bibinfo {author} {\bibfnamefont {O.}~\bibnamefont
  {Gomonay}}, \bibinfo {author} {\bibfnamefont {T.}~\bibnamefont {Jungwirth}},
  \ and\ \bibinfo {author} {\bibfnamefont {J.}~\bibnamefont {Sinova}},\ }\href
  {\doibase 10.1103/PhysRevLett.117.017202} {\bibfield  {journal} {\bibinfo
  {journal} {Phys. Rev. Lett.}\ }\textbf {\bibinfo {volume} {117}},\ \bibinfo
  {pages} {017202} (\bibinfo {year} {2016})}\BibitemShut {NoStop}%
\bibitem [{\citenamefont {Selzer}\ \emph {et~al.}(2016)\citenamefont {Selzer},
  \citenamefont {Atxitia}, \citenamefont {Ritzmann}, \citenamefont {Hinzke},\
  and\ \citenamefont {Nowak}}]{PhysRevLett.117.107201}%
  \BibitemOpen
  \bibfield  {author} {\bibinfo {author} {\bibfnamefont {S.}~\bibnamefont
  {Selzer}}, \bibinfo {author} {\bibfnamefont {U.}~\bibnamefont {Atxitia}},
  \bibinfo {author} {\bibfnamefont {U.}~\bibnamefont {Ritzmann}}, \bibinfo
  {author} {\bibfnamefont {D.}~\bibnamefont {Hinzke}}, \ and\ \bibinfo {author}
  {\bibfnamefont {U.}~\bibnamefont {Nowak}},\ }\href {\doibase
  10.1103/PhysRevLett.117.107201} {\bibfield  {journal} {\bibinfo  {journal}
  {Phys. Rev. Lett.}\ }\textbf {\bibinfo {volume} {117}},\ \bibinfo {pages}
  {107201} (\bibinfo {year} {2016})}\BibitemShut {NoStop}%
\bibitem [{\citenamefont {Yang}(2017)}]{yang_2017}%
  \BibitemOpen
  \bibfield  {author} {\bibinfo {author} {\bibfnamefont {L.}~\bibnamefont
  {Yang}},\ }\href {\doibase 10.4208/eajam.181016.300517d} {\bibfield
  {journal} {\bibinfo  {journal} {East Asian Journal on Applied Mathematics}\
  }\textbf {\bibinfo {volume} {7}},\ \bibinfo {pages} {837–851} (\bibinfo
  {year} {2017})}\BibitemShut {NoStop}%
\end{thebibliography}%

\end{document}